\documentclass[aps,prx,twocolumn,amsmath,amssymb,showpacs]{revtex4-1}

\usepackage[ansinew]{inputenc}
\usepackage{graphicx}
\usepackage{textcomp}
\usepackage{amsmath}
\usepackage{amssymb}
\usepackage{siunitx}
\usepackage{xcolor}
\usepackage{braket}

\begin{document}

\title{Coherent and Purcell-enhanced emission from erbium dopants in a cryogenic high-Q resonator}
\author{Benjamin Merkel}
\author{Alexander Ulanowski}
\author{Andreas Reiserer}
\email{andreas.reiserer@mpq.mpg.de}

\affiliation{Quantum Networks Group, Max-Planck-Institut f\"ur Quantenoptik, Hans-Kopfermann-Strasse 1, D-85748 Garching, Germany}
\affiliation{Munich Center for Quantum Science and Technology (MCQST), Ludwig-Maximilians-Universit\"at
M\"unchen, Fakult\"at f\"ur Physik, Schellingstr. 4, D-80799 M\"unchen, Germany}

\begin{abstract}
The stability and outstanding coherence of dopants and other atom-like defects in tailored host crystals make them a leading platform for the implementation of distributed quantum information processing and sensing in quantum networks. Albeit the required efficient light-matter coupling can be achieved via the integration into nanoscale resonators, in this approach the proximity of interfaces is detrimental to the coherence of even the least-sensitive emitters.

Here, we establish an alternative: By integrating a $19\,\si{\micro\metre}$ thin crystal into a cryogenic Fabry-Perot resonator with a quality factor of $9\cdot10^{6}$, we achieve a two-level Purcell factor of $530(50)$. In our specific system, erbium-doped yttrium orthosilicate, this leads to a $59(6)$-fold enhancement of the emission rate with an outcoupling efficiency of $46(8)\,\%$. At the same time, we demonstrate that the emitter properties are not degraded in our approach. We thus observe ensemble-averaged optical coherence up to $0.54(1)\,\si{\milli\second}$, which exceeds the $0.19(2)\,\si{\milli\second}$ lifetime of dopants at the cavity field maximum. While our approach is also applicable to other solid-state quantum emitters, such as color centers in diamond, our system emits at the minimal-loss wavelength of optical fibers and thus enables coherent and efficient nodes for long-distance quantum networks.
\end{abstract}
\maketitle

The implementation of large-scale quantum networks requires emitters with long ground-state coherence and coherent, spectrally indistinguishable optical transitions \cite{awschalom_quantum_2018}. In addition, scaling beyond the demonstrated two-node prototype networks requires highly efficient nodes \cite{wehner_quantum_2018}. In solid state realizations, achieving these properties simultaneously has proven difficult. Still, recent experiments suggest that this challenge can be overcome by embedding the emitters into an optical resonator \cite{sipahigil_integrated_2016, kindem_control_2020}. This reduces the emission on unwanted transitions and enables a large photon collection probability \cite{lodahl_interfacing_2015}, thus ensuring efficient quantum network nodes. Similarly, a large Purcell enhancement factor $P$ leads to spectral broadening that can overcome detrimental effects of spectral instability.

Achieving these benefits requires resonators with large quality factor $Q$ and small mode volume $V$, as $P\propto Q/V$. The latter has been achieved by confining light in nanophotonic structures \cite{lodahl_interfacing_2015, faraon_resonant_2011, sipahigil_integrated_2016, dibos_atomic_2018, kindem_control_2020, chen_parallel_2020}. Unfortunately, photon-mediated entanglement generation is impaired in nanostructured materials, which typically exhibit increased inhomogeneous broadening and thus spectral mismatch of embedded emitters. In addition, the proximity of interfaces causes a fluctuating charge and spin environment, which leads to reduced coherence and  diffusion of the optical transition frequencies \cite{lodahl_interfacing_2015, faraon_resonant_2011, sipahigil_integrated_2016, dibos_atomic_2018, kindem_control_2020, chen_parallel_2020}.

Therefore, good optical properties in nanophotonic structures have only been achieved with emitters that are insensitive to electric fields, such as silicon-vacancy centers in diamond \cite{sipahigil_integrated_2016} or rare-earth dopants in sites with low Stark coefficient \cite{kindem_control_2020}. Nanofabrication of suited resonators from these materials requires sophisticated techniques with limited yield and significant loss.

Here, we follow an alternative approach. Instead of using a nanostructured material, we embed a $19\,\si{\micro\meter}$ thin crystal slab in a cryogenic Fabry-Perot resonator. For ions at the slab center, this increases the distance to the next interface approximately 100-fold. One thus expects that the decoherence induced by surface charge and spin fluctuations is reduced by several orders of magnitude.

To still achieve a high Purcell enhancement, instead of minimizing $V$ we maximize $Q$ by using atomically flat membranes and low-loss dielectric coatings. We thus achieve a quality factor of $Q=9\cdot10^{6}$, about three orders of magnitude larger than state-of-the-art experiments with emitters in nanophotonic structures \cite{lodahl_interfacing_2015, faraon_resonant_2011, sipahigil_integrated_2016, dibos_atomic_2018,  kindem_control_2020, chen_parallel_2020}.

While our approach has been pioneered with cold atoms \cite{reiserer_cavity-based_2015}, cryogenic Fabry-Perot resonators with lower quality factor have recently been implemented with quantum dots \cite{najer_gated_2019}, thin diamond membranes \cite{riedel_deterministic_2017, bogdanovic_design_2017}, and rare-earth doped nanoparticles \cite{casabone_cavity-enhanced_2018, casabone_dynamic_2020}. Still, preserving the optical coherence of narrowband emitters while strongly enhancing their emission via the Purcell effect has not been demonstrated.

To this end, our experiments use erbium-dopants in yttrium-orthosilicate (YSO), whose transition wavelength around $1536.4\,\si{\nano\metre}$ ensures minimal loss in optical fibers, as required for quantum networks. Furthermore, the optical coherence of this transition is the best ever measured in a solid, up to $4\,\si{\milli\second}$ \cite{bottger_effects_2009}. While the resulting exceptionally narrow linewidth might offer unique potential for spectrally multiplexed spin-qubit readout \cite{chen_parallel_2020}, it has not been preserved upon integration into nanostructured resonators \cite{miyazono_coupling_2016, dibos_atomic_2018}. In contrast, our approach allows for coherence exceeding the lifetime of the strongest-coupled dopants, which is a key enabling step for remote entanglement generation \cite{awschalom_quantum_2018}.

\begin{figure}
\includegraphics[width=1.0\columnwidth]{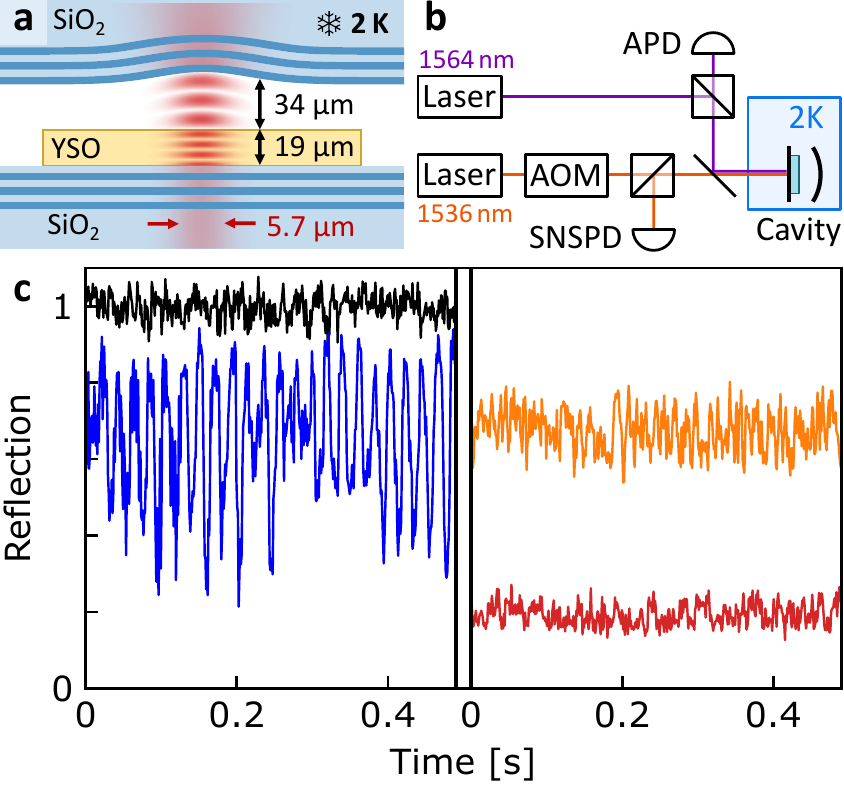}

\caption{\label{fig:setup}
\textbf{Experimental setup.} \textbf{a. Fabry-Perot resonator} (not to scale). A stable cavity mode (red) with $\sim 46(8)\,\%$ outcoupling efficiency towards one side is obtained by embedding a $19(1)\,\si{\micro\metre}$ thin membrane of erbium-doped YSO (orange) between two dielectric mirrors (blue layers of alternating refractive index). \textbf{b. Optical setup.} The cavity is mounted at $2\,\si{\kelvin}$ in a closed-cycle cryocooler (blue box). It is probed with a faint laser field, polarized along the D1 axis of YSO, and on resonance with the transition of erbium dopants around $1536.4\,\si{\nano\metre}$. It is switched by an acousto-optical modulator (AOM) and detected by a superconducting single-photon detector (SNSPD). A second laser field at $1564\,\si{\nano\metre}$ is resonant with another longitudinal resonator mode and can be used for resonator stabilization. \textbf{c. Resonator stability over time}. Off-resonant light is fully reflected from the cavity (black). Without active stabilization (left panel), vibrations lead to a fluctuating signal when tuned on resonance (blue). With thermooptical feedback (right), the cavity frequency can be stabilized to the side (orange) or bottom (red) of the reflection.
}
\end{figure}

A sketch of our experimental system is shown in Fig.~\ref{fig:setup}. We use a plano-concave Fabry-Perot resonator with a tunable length around $L=50\,\si{\micro\metre}$ and a mirror radius of curvature $R=155\,\si{\micro\metre}$, giving a $w_0=5.7\,\si{\micro\metre}$ waist of the fundamental Gaussian mode. In the cavity, an atomically-flat, $19(1)\,\si{\micro\metre}$ thin YSO membrane serves as host for erbium dopants with an estimated concentration $<0.3\,\text{ppm}$ (see supplemental material). The mirror transmissions of $7(1)$ and $24(1)$ ppm for the concave and crystal-covered side are comparable to the absorption and scattering losses, $21(8)$ ppm, leading to a finesse of $1.2(2)\cdot 10^5$, a linewidth of $22(2)\,\si{\mega\hertz}$, and an outcoupling efficiency of $\sim 46(8)\,\%$ to a single-mode fiber. 

Keeping the cavity resonant requires to control the mirror separation to less than a picometer. Similar to other experiments with lower finesse \cite{bogdanovic_design_2017, casabone_cavity-enhanced_2018, casabone_dynamic_2020}, this turned out challenging in our closed-cycle cryostat. We therefore combine a vibration isolation platform with fine tuning via a piezo tube. Still, we observe fluctuations of the cavity resonance frequency with a root-mean-square of $8(2)\,\si{\mega\hertz}$, comparable to the cavity linewidth, as shown in Fig.~\ref{fig:setup}c. To eliminate the residual vibrations via photothermal feedback, we irradiate a laser field at a wavelength of $1564\,\si{\nano\metre}$, far detuned from the erbium transition but close to resonance with another longitudinal cavity mode. When the resonator fluctuates, a higher or lower fraction of the laser light is absorbed, which changes the crystal temperature and thus shifts the resonance. Choosing suited parameters, we can thus stabilize the resonator frequency \cite{brachmann_photothermal_2016}.

Before characterizing the effects of this temperature change, we investigate the optical properties of the erbium ensemble by measuring fluorescence after resonant pulsed excitation. At temperatures around $2\,\si{\kelvin}$, we apply a magnetic field of $0.8\,\si{\tesla}$ along the D2 axis of the crystal to ensure that the population is trapped in the probed lower spin state, avoiding detrimental effects of spectral hole-burning \cite{car_optical_2019} and superhyperfine couplings \cite{car_superhyperfine_2020}.

We first investigate the inhomogeneous broadening by sweeping the laser and cavity frequency. In contrast to nanofabricated resonators on the same host \cite{dibos_atomic_2018, chen_parallel_2020}, the observed Lorentzian spectrum with a FWHM of $414(7)\,\si{\mega\hertz}$ (see supplemental material) is not broadened compared to bulk reference crystals. This testifies low strain in our crystalline membrane and indicates that our dopant concentration is low enough to avoid effects of collective strong coupling.

\begin{figure}
\includegraphics[width=1.0\columnwidth]{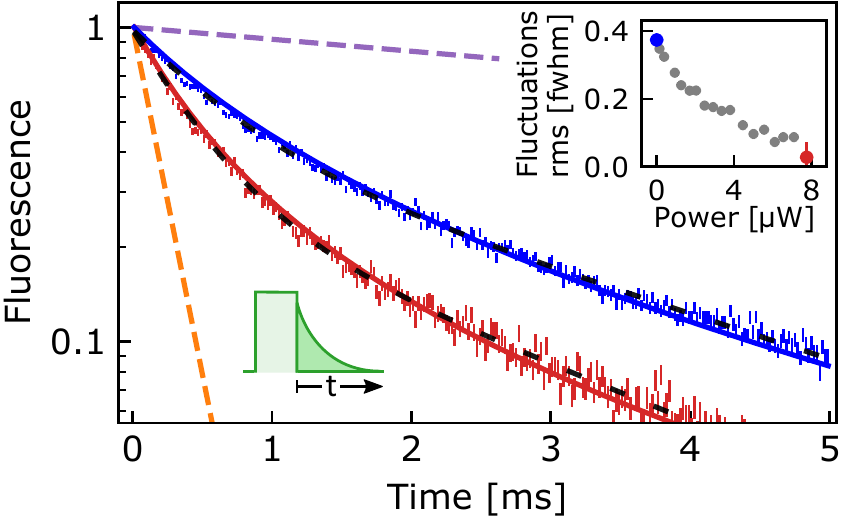}

\caption{\label{fig:PurcellEnhancement}
\textbf{Purcell-enhanced emission}. After pulsed excitation with a power of $54\,\si{\nano\watt}$ and a duration of $0.5\,\si{\micro\second}$, the fluorescence of resonator-coupled dopants (blue) decays on a faster timescale than in free-space (purple dashed line). The decay rate is further enhanced when the cavity is kept on resonance via photothermal feedback (red). In both cases, the decay is well fit by a biexponential curve (black dashed), and in excellent agreement with a numerical modeling of the resonator with no free parameters (red and blue solid curves). According to this model, the decay of dopants at the field maximum (orange dashed line) corresponds to a lifetime-reduction factor of $59(6)$. The inset shows the fluctuations of the cavity length as a function of feedback laser power, derived by fitting the measured decay to our model. 
}
\end{figure}

As a next step, we measure the fluorescence at the center of the inhomogeneous distribution. As the dopants are randomly distributed across the Gaussian standing-wave mode of the resonator, the observed signal is an average over many Purcell-enhanced decays with different lifetime. As shown in Fig.~\ref{fig:PurcellEnhancement}, our data is well fit by a biexponential decay. A longer lifetime is observed when the cavity is vibrating and thus spends less time on resonance with the dopants.

To gain further insight into the observed decay, we calculate the expected fluorescence using the independently determined cavity properties \cite{janitz_fabry-perot_2015}, as detailed in the supplemental material. This gives a large Purcell enhancement of $P_\text{TL}=530(50)$ for a two-level system at the maximum of the cavity field. However, the $11\,\%$ branching ratio of the investigated transition \cite{miyazono_coupling_2016}, calculated from  absorption and lifetime measurements \cite{bottger_spectroscopy_2006}, leads to a lower value. For maximally coupled dopants, we thus expect a lifetime reduction by a factor of $P_\text{Er,max}+1=59(6)$, which gives $T_1=0.19(2)\,\si{\milli\second}$.

As detailed in the supplemental material and shown in Fig. \ref{fig:PurcellEnhancement}, when scaling the calculated Purcell factor with the electric field intensity and averaging over a random dopant distribution in the resonator, the obtained theoretically expected fluorescence curve is in excellent agreement with the data (with no free parameters). 

The lifetime is unchanged when the excitation pulse power and thus the number of excited dopants is increased over several orders of magnitude (as shown in the supplemental material). We can thus preclude that cavity-enhanced superradiance contributes to the observed lifetime reduction.

\begin{figure}
\includegraphics[width=1.0\columnwidth]{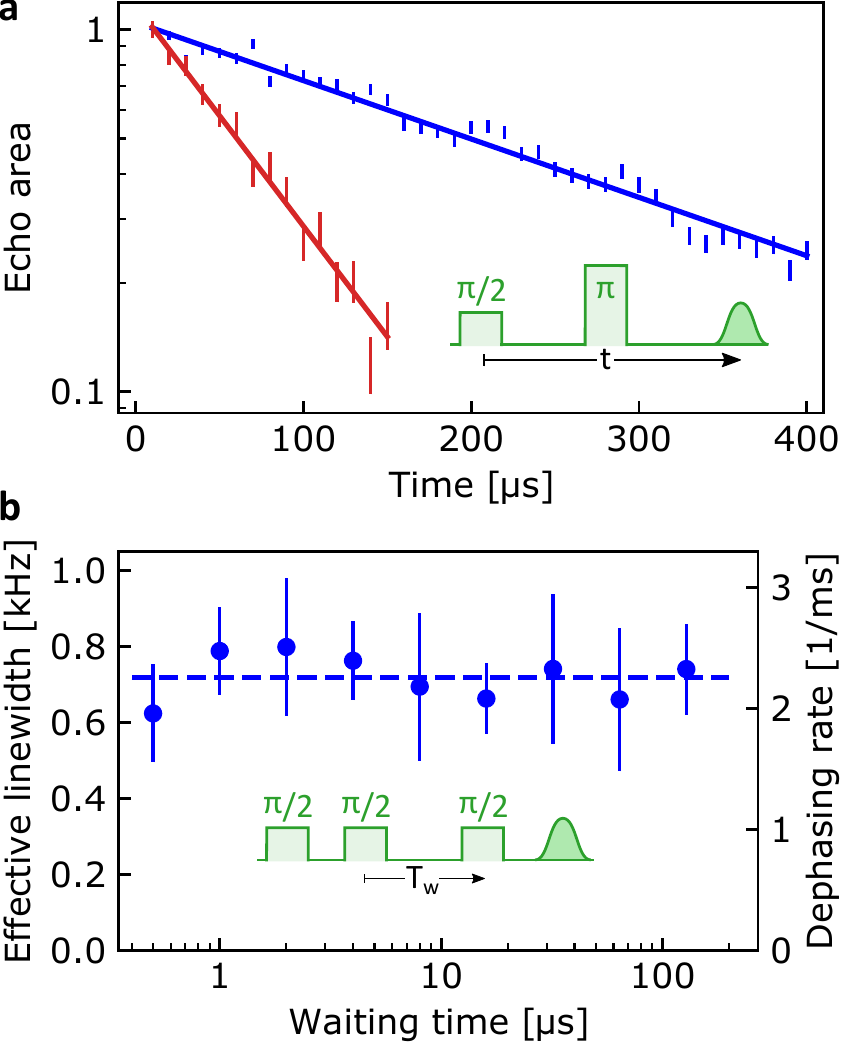}

\caption{\label{fig:optical_coherence}
\textbf{Optical coherence measurement. a. Two-pulse photon echo.} The used sequence,  with $0.5\,\si{\micro\second}$ long optical $\pi$ pulses with a power of $54\,\si{\nano\watt}$, is shown in the inset. As expected in the absence of spectral diffusion during the optical lifetime, the coherence decays exponentially with the echo time $t$. Compared to the situation without photothermal feedback (blue), a faster decay is observed when the resonator frequency is stabilized (red), which we attribute to heating of the crystal. \textbf{b. Three-pulse photon echo.} We measure the echo after three $\pi/2$ pulses, as shown in the inset. We observe a constant effective linewidth $\Gamma_\text{eff}$ independent of the waiting time $T_w$, which further testifies the absence of spectral diffusion within the $0.19(2)\,\si{\milli\second}$ Purcell-enhanced lifetime of the strongest-coupled dopants.
}
\end{figure}

We now turn to the optical coherence of the dopants which we study via photon-echo measurements \cite{bottger_effects_2009, car_superhyperfine_2020}. The used pulse sequence is shown as an inset in Fig.~\ref{fig:optical_coherence}a. A first optical pulse with an area of approximately $\pi/2$ prepares an optical coherence that quickly dephases because of the inhomogeneous broadening of our emitters. A second optical $\pi$-pulse, irradiated after a time $t/2$, leads to a rephasing and thus to the emission of an echo signal at time $t$. Fig.~\ref{fig:optical_coherence}a shows the measured echo area as a function of $t$ both with (red) and without (blue) photothermal feedback. An exponential fit leads to an optical coherence of $T_2=0.14(1)\,\si{\milli\second}$ and $T_2=0.54(1)\,\si{\milli\second}$, respectively.

The observed coherence of the probed ensemble thus exceeds the lifetime of dopants at the field maximum, $T_2>T_1$, achieving a key requirement for efficient remote entanglement generation without excessive temporal filtering \cite{nolleke_efficient_2013, reiserer_cavity-based_2015}. 

Remarkably, we even observe a tenfold improvement of the coherence compared to recent measurements with bulk crystals \cite{car_superhyperfine_2020}, which is explained by the reduction of dopant interactions \cite{merkel_dynamical_2020, car_optical_2019} at our lower concentration. In addition, as in both measurements the decay is well fit by a single exponential, we have no indication of homogeneous line broadening via spectral diffusion \cite{bottger_optical_2006} for maximally Purcell-enhanced dopants.

To confirm this, we performed three-pulse echo measurements, as shown in Fig.~\ref{fig:optical_coherence}b. Here, for several values of the waiting time $T_w$ we determine the effective linewidth by fitting the exponential decay of the echo area when varying the separation between the first two pulses. The observation of a constant effective linewidth up to $T_w=0.2\,\si{\milli\second}$ indicates that the amplitude or the rate of spectral diffusion in our system is too low to cause decoherence during the lifetime of the best-coupled emitters \cite{liu_spectroscopic_2005}.

In the present configuration, we expect spectral diffusion processes with an effective linewidth of a few hundred kHz on longer timescales, presumably seconds, caused by dipolar interactions with the fluctuating nuclear spin bath \cite{bottger_optical_2006}. Potential detrimental effects can be overcome by a suited choice of the magnetic field direction that gives a zero first-order Zeeman shift, or by regular measurements and compensation of resonance frequency changes \cite{robledo_control_2010}.

As mentioned, we observe a faster decay of the coherence in the presence of photothermal feedback. We attribute this to the increased temperature. To study this effect, we first confirm that the crystal is thermalized with an adjacent thermometer by comparing the population of the two ground-state spin levels.

\begin{figure}
\includegraphics[width=1.0\columnwidth]{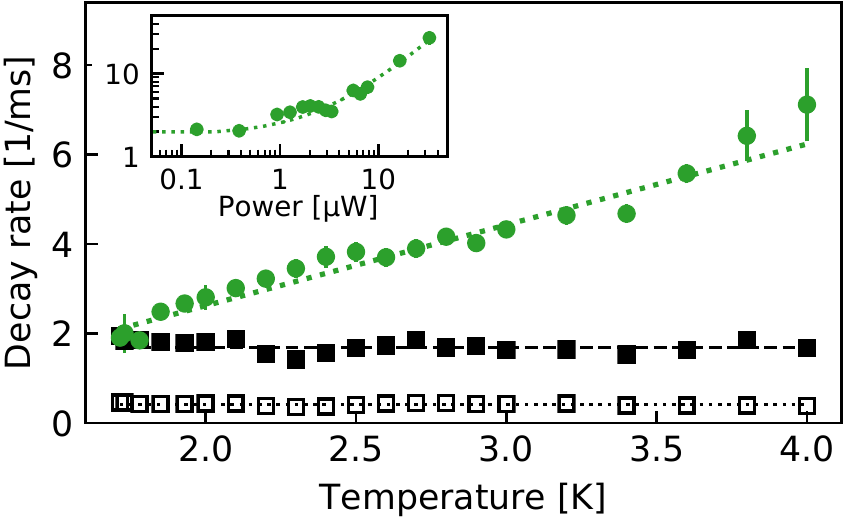}

\caption{\label{fig:T_dependence}
\textbf{Temperature dependence}. When the temperature of the sample and exchange gas are increased, the fast (black filled) and slow (black open rectangles) decay constants of the fluorescence remain unchanged (black fit lines). In contrast, the decay rate of the coherence increases (green data and linear fit). The inset shows the rise of the decay rate with increased thermooptical feedback laser power (green data and linear fit). Based on these measurements, we estimate that at the current vibration level, complete stabilization of the resonator at $\approx 5\,\si{\micro\watt}$ (c.f. Fig.~\ref{fig:PurcellEnhancement}) requires a temperature increase of $\sim 2\,\si{\kelvin}$.
}
\end{figure}

We then increase the crystal temperature using a resistive heater. As shown in Fig.~\ref{fig:T_dependence}, the decay rate of the coherence, $1/T_2$, rises with temperature (green), while the lifetime remains unchanged (black). Similarly, the dephasing is increased linearly with the power of the photothermal feedback laser (inset). 

At the power required to stabilize the resonator frequency to $1/10$ of its linewidth at the given mechanical fluctuations, $\sim5\,\si{\micro\watt}$, the dephasing rate is $\sim 5\,\si{ms^{-1}}$. Comparing the curves, we conclude that in this case the crystal temperature is increased by $\sim 2\,\si{\kelvin}$. We expect that such temperature increase can be tolerated in other physical platforms that are less affected by thermal decoherence, in particular the NV center in diamond. Here, the demonstrated photothermal feedback can be readily used to suppress the detrimental effects of vibrations that have hampered the use of cryocoolers \cite{bogdanovic_design_2017}.

In summary, we have shown that integration into high-finesse Fabry-Perot resonators allows for a strong increase of the coupling of dopants to light while preserving their outstanding optical coherence properties.

Our system  thus enables efficient, cavity-enhanced quantum memories \cite{afzelius_quantum_2015}, in which ultra-low dopant concentration overcomes the limitations imposed by spin interactions \cite{merkel_dynamical_2020}. In addition, the possibly good mode overlap with superconducting resonators facilitates efficient microwave-to-optical conversion \cite{williamson_magneto-optic_2014, chen_coupling_2016}. Furthermore, spectral holeburning at higher dopant concentration \cite{krimer_hybrid_2015} might be used for laser stabilization \cite{cook_laser-frequency_2015} in the telecom C-band for fiber-based sensing or optical atomic clocks.

While the demonstrated possibility to largely decouple the vibrations of a closed-cycle cryocooler may find applications in optomechanical systems \cite{aspelmeyer_cavity_2014}, it will also give a boost to quantum networking experiments with host crystals that are not amenable to nanofabrication \cite{awschalom_quantum_2018}. Using  fiber-coupled \cite{najer_gated_2019, bogdanovic_design_2017,casabone_cavity-enhanced_2018, casabone_dynamic_2020} or chip-based \cite{wachter_silicon_2019} Fabry-Perot resonators, we still expect that scalable fabrication of quantum network nodes will be feasible.

In our system, the Purcell enhancement can be further increased. First, aligning the polarization along the D2 axis of YSO \cite{miyazono_coupling_2016} increases the dipole moment twofold compared to our measurements. Second, a tenfold enhancement is expected for mirrors with smaller radius of curvature \cite{reiserer_cavity-based_2015, najer_gated_2019, bogdanovic_design_2017,casabone_cavity-enhanced_2018, wachter_silicon_2019, casabone_dynamic_2020}. Finally, the achieved surface quality allows for increasing the finesse by another order of magnitude with better mirror coatings \cite{reiserer_cavity-based_2015, najer_gated_2019,  wachter_silicon_2019}.

Even without these improvements, by operating at the tail of the inhomogeneous broadening \cite{kindem_control_2020} we plan to resolve and control single erbium dopants with good optical coherence. Alternatively, host crystals without rare-earth contamination \cite{bertaina_rare-earth_2007, phenicie_narrow_2019, weiss_erbium_2020}, or crystals with larger inhomogeneous broadening can be used to this end. Combined with frequency-multiplexed readout of individual spins \cite{chen_parallel_2020} and the long ground state coherence of the $^{167}\text{Er}$ nuclear spin \cite{rancic_coherence_2018}, our approach may thus facilitate quantum network nodes operating at $^4\text{He}$ temperature and at telecommunication wavelength. 

\section*{Acknowledgements}
This project received funding from the European Research Council (ERC) under the European Union's Horizon 2020 research and innovation programme (grant agreement No 757772), from the Deutsche Forschungsgemeinschaft (DFG, German Research Foundation) under Germany's Excellence Strategy - EXC-2111 - 390814868, and from the Daimler-and-Benz-Foundation. We acknowledge the contribution of Natalie Wilson, Manuel Brekenfeld, and Dominik Niemietz in fabricating the concave mirror, and discussions with Alexander Kubanek.
\newpage
\section*{Supplementary information}
\subsection{Experimental setup}

The sample is an undoped YSO crystal grown by Scientific Materials and cut along the polarization axes $D_1$, $D_2$ and $b$. The laser light propagates along the $b$ axis with a polarization $\parallel D_1$. The magnetic field is applied $\parallel D_2$. Trace impurities lead to an erbium concentration of $<0.3\,\text{ppm}$, estimated from the saturated fluorescence level and the known resonator mode volume. To tune the cavity length, the mirrors are attached to a custom-made piezo tube (PI ceramics) via titanium springs. Details about the sample and resonator fabrication are described in the next section.

The cavity is mounted in a closed-cycle cryostat designed for low vibrations and variable temperature down to $1.6\,\si{\kelvin}$ (AttoDry 2100). To achieve sufficient passive stability of the mirror distance, we hang the resonator and coupling optics with glass fiber rods from a home-built vibration isolation platform based on passive air-dampers with active position control (Thorlabs PWA090) on top of the cryostat at ambient temperature. Using a flexible edge-welded bellow filled with helium, full thermalization of the sample with its surrounding tube is achieved down to temperatures of $1.72\,\si{\kelvin}$. This is confirmed by resonantly exciting the fluorescence of the dopants in both spin ground states under a magnetic field of $0.8\,\si{\tesla}$ generated by a superconducting solenoid. Assuming a Boltzmann distribution, the ratio of the measured fluorescence is in good agreement with the reading of a resistive thermometer in close proximity to the crystal.

While we reached sub-pm stability in a previous measurement, the performance of the vibration isolation system has been degraded after replacing the variable-temperature insert of the cryostat following fatal damage caused by a short contact. Therefore, all data in this study is acquired under worse mechanical stability. A possible explanation is that the resonator is touching the vacuum tube side wall, which reduces the efficiency of the vibration isolation. This can be fixed in a future redesign with increased mechanical tolerances.

\begin{figure}
\includegraphics[width=1.0\columnwidth]{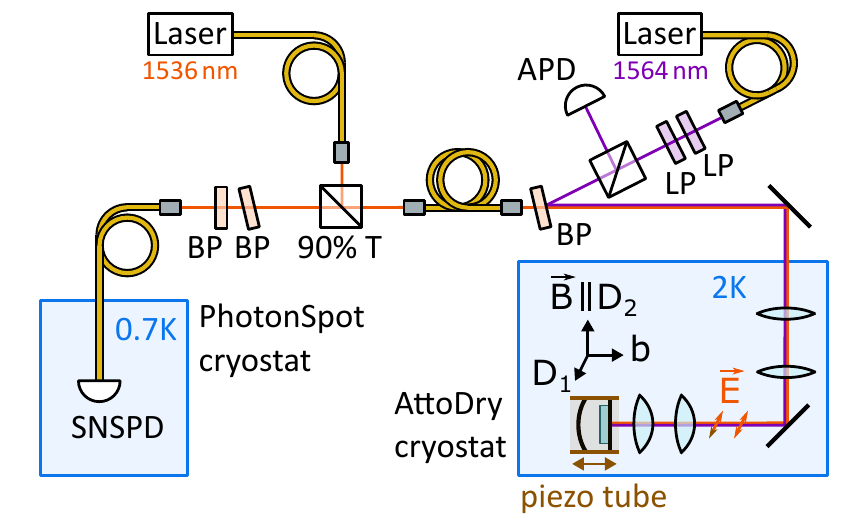}
\caption{\label{fig:supsetup}
\textbf{Detailed experimental setup.} A frequency-stabilized, off-resonant laser beam (purple) at a wavelength of $1564\,\si{\nano\metre}$ is used to determine the resonator length and stabilize it on both short and long ($\sim\,\si{\milli\second}$ to several days) timescales via thermo-optical feedback and a piezo tube, respectively. It first passes two filters (LP) to suppress amplified spontaneous emission, before it is combined with the resonant probe laser (orange) at a band-pass filter (BP) under a slight angle that acts as a polarization-insensitive dichroic mirror. Both lasers are coupled into the resonator via a 1:1 telescope and a custom designed combination of aspheric lenses that ensure good mode matching. The reflected light is separated from the input by 90:10 beam splitters and detected with either an avalanche photodiode (APD) or a superconducting nanowire single photon detector (SNSPD) in a second cryostat.
}
\end{figure}

As shown in Fig. \ref{fig:supsetup}, for thermo-optical feedback and fluorescence excitation we use tunable diode laser laser systems (Toptica CTL 1550 and DLpro as well as OEwaves Gen3). The lasers are stablized to a frequency comb (Menlo Systems) and switched via acousto-optical modulators (Gooch and Housego). The light is detected with an avalanche photodiode (Thorlabs) or a superconducting nanowire single photon detector (Photon Spot) after suited spectral filters (Semrock). The experiment is controlled and the data is recorded by a real-time experimental control system (National Instruments Compact RIO).

\subsection{Crystal and resonator fabrication}

The cavity consists of two highly-reflective mirrors. Their fabrication starts from superpolished substrates with $7\,\si{\milli\metre}$ diameter (Research Electro Optics Inc.). One of the substrates is machined with a $\text{CO}_2$ laser at a wavelength of $9.3\,\si{\micro\metre}$ to fabricate concave depressions using a setup described in \cite{uphoff_frequency_2015}. We create a regular 4 by 4 array of nearly Gaussian depressions, with a depth around $20\,\si{\micro\metre}$ and effective radii of curvature between $150$ and $250\,\si{\micro\metre}$.

Both mirrors are coated (Laseroptik GmbH) with a dielectric stack of alternating $\text{SiO}_2$ and $\text{Ta}_2\text{O}_5$ layers. The coating of the curved mirror is terminated by $\text{Ta}_2\text{O}_5$ and has a design transmission of $10\,\text{ppm}$, while the flat mirror is terminated by $\text{SiO}_2$ with a design transmission of $100\,\text{ppm}$.

The mirrors are held at a stable distance by titanium springs that are attached to a piezo tube (PI ceramics). At room temperature, this allows for scanning the mirror distance over more than 10 free spectral ranges of the resonator, which reduces to less than one upon cryogenic operation.

Between the mirrors, a $19(1)\,\si{\micro\metre}$ thin membrane of YSO is used as a host for erbium dopants. While this host is well known to allow for excellent coherence properties, fabrication of membranes turned out to be challenging. We first tried reactive-ion etching with oxygen, argon and fluoride gases in different compositions. All used recipes led to a slow etch rate and increased surface roughness or surface amorphization, both of which prevent integration into a high-finesse resonator. We therefore developed a chemo-mechanical polishing procedure together with the crystal laboratory at TU Munich and Optec Munich. We now reproducibly achieve a surface roughness below $0.3\,\si{\nano\meter}$ for membranes down to $10\,\si{\micro\metre}$ thickness.

The procedure starts by polishing one side of the crystal to the desired surface quality. Then, this side is glued to a flat glass substrate using a thin film of low-viscosity glue. After curing, the second side of the crystal is polished until the desired thickness is achieved.

To transfer the crystal to a cavity mirror without introducing excess strain, we start by gently cleaning the crystal with isopropanol and acetone. We then cover a glass ``carrier'' with teflon tape and add a drop of acetone before putting it into a clean vessel. Next, we flip the glass substrate with the crystal onto the acetone drop on the carrier, such that it sticks via surface tension. We fill the vessel with acetone and wait until the glue has dissolved and all acetone has evaporated.

When the glass substrate is lifted, the crystal is left on the dry PTFE coated carrier. Still, both sides of the crystal show residues of the glue. To remove them, we add a little drop of acetone to the edge such that it creeps below the crystal, making it stick to the teflon because of surface tension. We can then gently and repeatedly clean the top surface with acetone- and isopropanol-soaked lens paper, removing residues of the glue from the top.

After getting a clean surface in a white-light microscope image, we wait until the acetone below the membrane has evaporated. We then press the mirror to the crystal, where it sticks via van-der-Waals forces. After flipping, we again gently clean the crystal surface with isopropanol-soaked lens paper until all residues of the glue are removed. This procedure works reproducibly for 5 by $5.5\,\si{\milli\metre}^2$ crystalline membranes.

Fig.~\ref{fig:crystalPrep}a shows the center part of the crystal surface after transfer, where the height profile has been measured by a white-light interferometer. Except for a few dust particles with heights of several $\si{\nano\meter}$, a very flat surface is achieved.

Fig.~\ref{fig:crystalPrep}b shows two line cuts at the center along the x and y direction. Over the $\simeq 10\,\si{\micro\meter}$ mode diameter of our cavity, a root-mean-square surface roughness below $0.2\,\si{\nano\metre}$, likely limited by the noise of the instrument, is achieved. We thus expect that the scattering of the crystal does not limit the finesse of a surrounding resonator even for values exceeding $10^6$ \cite{hunger_laser_2012}.

\begin{figure}
\includegraphics[width=1.0\columnwidth]{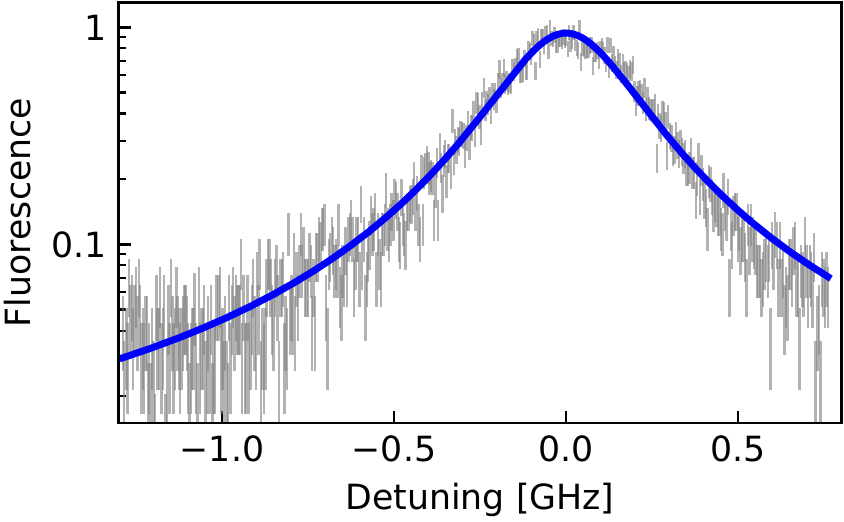}
\caption{\label{fig:inhombroadening}
\textbf{Inhomogeneous broadening.} The laser and cavity are swept around the frequency of the erbium transition while recording pulsed fluorescence. The data was taken at a magnetic field of $3\,\si{\tesla}$ for a single magnetic class and is well fit by a Lorentzian curve (blue) with a FWHM of $414(7)\,\si{\mega\hertz}$.
}
\end{figure}

To measure the strain in the membrane at cryogenic temperature, we compare the inhomogeneous broadening of the membrane in the resonator to that of several reference crystals grown by the same company with the lowest available dopant concentration of $10\,\text{ppm}$. We find a shift of $\sim 1.5\,\si{\giga\hertz}$, but no additional broadening, as shown in Fig.~\ref{fig:inhombroadening}. This indicates that in spite of the mismatch in thermal expansion coefficients of $\text{SiO}_2$ and YSO, no significant strain gradient is generated in the membrane upon cooldown.

\begin{figure*}
\includegraphics[width=2.0\columnwidth]{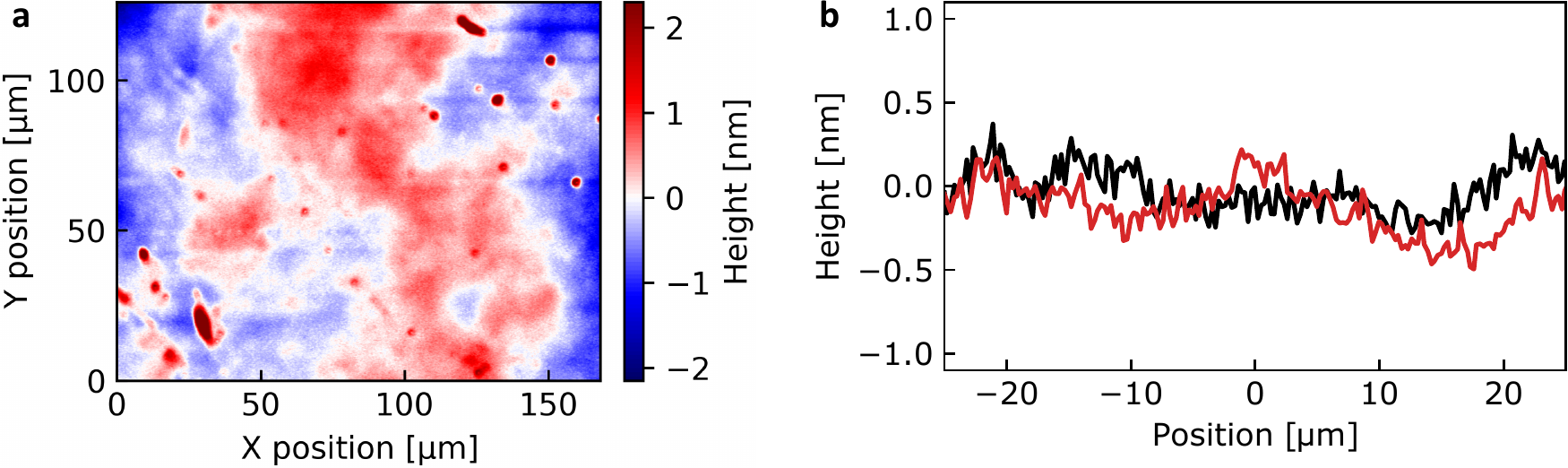}
\caption{\label{fig:crystalPrep}
\textbf{Crystal properties. a.} Surface height profile of the YSO membrane after transfer to the mirror. At its center, the membrane is fully bonded to the mirror and shows excellent flatness. Apart from a few dust particles that might be removed by further cleaning (red dots), the surface height variation is well below $1\,\si{\nano\meter}$.
\textbf{b.} As can be seen from line cuts at the center, the surface roughness is below $0.2\,\si{\nano\metre}$ over the $\simeq10\,\si{\micro\meter}$ mode diameter of our resonator.
}
\end{figure*}

\subsection{Cavity parameters}
To model the Purcell enhancement in the assembled Fabry-Perot resonator, we start by analyzing the electromagnetic field using the methods described in \cite{janitz_fabry-perot_2015}. To this end, we irradiate a laser (Toptica CTL 1550) and measure its transmission with a photodiode or infrared camera (NIT WiDy SWIR 320) to determine the transversal mode. The birefringence of the host crystal clearly separates the resonance frequency for orthogonal polarization. For each polarization, we observe a set of Hermite-Gaussian modes, where the higher-order modes are slightly split because of a small ellipticity of the concave cavity mirror \cite{uphoff_frequency_2015}.

We scan the laser from 1520 to 1630 nm and determine the frequency of all resonant modes with a calibrated wavemeter (Bristol 771 series OSA). Thus, the free-spectral range and higher-order mode splitting is determined. Including a phase correction term \cite{garmire_theory_2003} in the modeling of Ref.~\cite{janitz_fabry-perot_2015} allows us to calculate the crystal thickness $t_c=18.2(1)\,\si{\micro\metre}$, air gap $t_a=29.8(1)\,\si{\micro\metre}$, and radius of curvature $R=155(3)\,\si{\micro\metre}$ of the concave mirror. To this end, we implement a least-square fitting procedure that combines a grid search with a local optimization algorithm.

We then continue to determine the losses of the resonator. To this end, we first measure its linewidth. We modulate the laser with an electro-optical modulator, which creates two sidebands symmetrical around the cavity resonance. We then apply a linear frequency ramp to the laser and use the sidebands to gauge the frequency axis. The linewidth is then obtained by Lorentzian fits of the cavity transmission. To reduce inaccuracy that may arise from mechanical fluctuations, the fit results of many runs are averaged.

\begin{figure*}
\includegraphics[width=2.0\columnwidth]{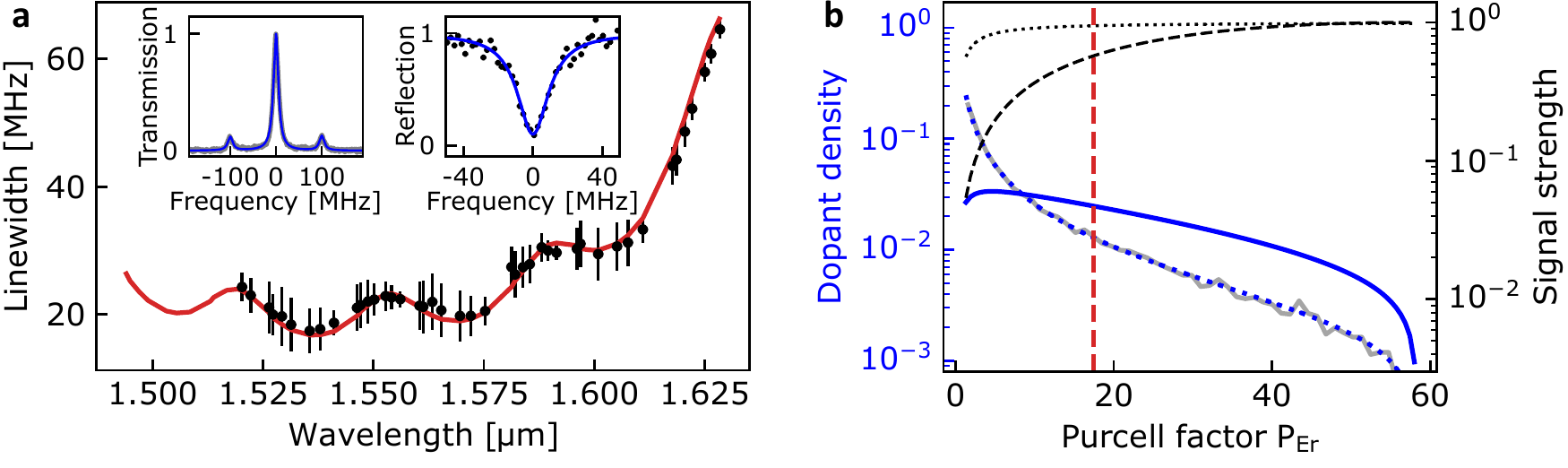}
\caption{\label{fig:cavity model}
\textbf{Modeling of the cavity. a.} The cavity resonance can be measured in transmission (left inset, not stabilized, with laser modulation) and reflection (right inset, with photothermal feedback, without modulation). On resonance, the reflection drops close to zero, which indicates that the resonator is almost critically coupled and the mode-matching of the cavity fundamental mode to a light field propagating in an optical fiber is excellent.
The fitted linewidth changes as a function of wavelength (main graph, here at room temperature), which can be accurately modeled over the entire range (red) using a transfer-matrix-approach \cite{janitz_fabry-perot_2015} that includes the crystal and dielectric layers of the mirror coatings. This allows us to determine the crystal thickness and parameters of the assembled cavity.
\textbf{b.} The intensity profile of the cavity mode leads to a probability distribution of dopants with different Purcell factors $p(P_\text{Er})$ and can be extracted from either a Monte Carlo simulation of 30000 dopants (solid grey curve) or eq.~\ref{eq:p_Er} (blue dotted curve). Since the probabilities of exciting a dopant (black dashed curve) and emitting a photon into the cavity mode (black dotted curve) reduce the fluorescence signal caused by weakly-coupled dopants, the effective distribution of Purcell factors (solid blue curve) contributing to the fluorescence decay signal has an average value of $P_\text{Er,avg}=17$ (red dashed line). 
}
\end{figure*}

In our hybrid cavity, the linewidth depends on the resonance frequency \cite{janitz_fabry-perot_2015}, as can be seen in Fig.~\ref{fig:cavity model}a. For the Erbium transition frequency, we obtain a linewidth of $22(2)\,\si{\mega\hertz}$ FWHM, which gives a field decay rate $\kappa = 2\pi \cdot 11(1)\,\si{\mega\hertz}$. In addition to the total loss, we are also interested in the photon outcoupling rates through the mirror on the crystal-side of the cavity and the mirror on the air-side. These can be determined by measuring the transmission and reflection of the system at room temperature from both sides, which eliminates the unknown coupling efficiency of the impinging laser field. Upon cooldown, the cavity length is reduced by $5\,\si{\micro\meter}$. We then find $\kappa_\text{crystal} = 2\pi \cdot 5(1)\,\si{\mega\hertz}$ and $\kappa_\text{air} = 2\pi \cdot 1.5(3)\,\si{\mega\hertz}$, which gives a total outcoupling efficiency of $(\kappa_\text{crystal}+\kappa_\text{air})/\kappa=60\,\%$.

Furthermore, the polarization decay rate $\gamma=1/(2T_0)$ can be calculated from the radiative lifetime and absorption coefficient measured in weakly doped crystals at the same orientation \cite{bottger_optical_2006}. In contrast to recent experiments with nanophotonic structures \cite{miyazono_coupling_2016}, we are using a bulk crystal. Therefore, no additional local-field corrections are required in our modeling.

In summary, we can give the complete set of parameters that describe the dynamics of maximally coupled dopants in our cavity:
\begin{align*}
    g_\text{max} &= 2\pi \cdot 67(7)\,\si{\kilo\hertz} \\
    \kappa &= 2\pi \cdot 11(1)\,\si{\mega\hertz} \\
    \gamma &= 2\pi \cdot 7\,\si{\hertz}
\end{align*}

Here, the maximum coupling strength $g_\text{max}$ has been calculated from the cavity geometry with mode volume $V=750(10)\,\si{\micro\metre}^3$ and the oscillator strength of the used transition $1\cdot 10^{-7}$ \cite{bottger_spectroscopy_2006}. As mentioned in the main text, the resulting maximum Purcell factor for an Erbium dopant, $P_\text{Er,max}=g_\text{max}^2/(\kappa\gamma)=58(6)$, can also be calculated from the maximum Purcell factor for a two-level system, $P_\text{TL}=3\lambda^3Q/(4\pi^2n^3 V)=530(50)$, and the branching ratio of $11\,\%$ extracted in \cite{miyazono_coupling_2016}, arriving at the same result. 

\subsection{Numerical modeling of the fluorescence measurement}
In the preceding section, we have determined all relevant cavity QED parameters of our system, which gives a Purcell enhancement of $P_\text{Er,max}=58(6)$ for an emitter at the maximum of the mode. Most Erbium dopants, however, are not located at a field maximum and experience a reduced Purcell factor, $P_\text{Er}$. In this section, we explain how we calculate the expected fluorescence decay from the electromagnetic field profile of the cavity mode within the crystal. 

We expect the envelope of the electromagnetic field inside the crystal to follow a standing wave pattern along the resonator axis $z$ and a Gaussian profile of the fundamental mode along the radial direction $\rho$:
\begin{equation}
E(\rho,z)= E_\text{max} \cdot\sin(2\pi n z/\lambda) \cdot \exp(-\rho^2/w_0^2)
\end{equation}
Here, $n$ is the refractive index of the crystal, $\lambda$ the vacuum wavelength, and $w_0$ the mode waist. To calculate the Purcell factor at any position ($\rho, z$), we take the value $P_\text{Er,max}$ for emitters at the field maximum, $E_\text{max}$, and scale it with the square of the field envelope (consisting of the axial component $u_z(z)$ and the radial component $u_\rho(\rho)$):
\begin{equation}\label{eq:P_Er(rho,z)}
P_\text{Er}(\rho, z)= P_\text{Er,max} \cdot \underbrace{\sin^2(2\pi n z/\lambda)}_{u_z(z)} \cdot \underbrace{\exp(-2\rho^2/w_0^2)}_{u_\rho(\rho)}
\end{equation}

To obtain the distribution of Purcell factors in our ensemble of cavity-coupled dopants, one can make a Monte Carlo simulation of homogeneously distributed emitters and calculate their individual Purcell factors according to eq.~\ref{eq:P_Er(rho,z)} (solid grey line in Fig.~\ref{fig:cavity model}b). 

It is, however, also possible to calculate an analytical expression for the probability density $p(P_\text{Er})$ to find an Erbium dopant with Purcell factor $P_\text{Er}$, which we will sketch in the following. As a starting point, the probability density $p(x,y,z)$ to find an emitter at Cartesian coordinates $x,y,z$ is constant, assuming a homogeneous distribution. According to eq.~\ref{eq:P_Er(rho,z)}, the axial and radial intensity envelopes $u_z$ and $u_\rho$ can be calculated separately, which allows us to derive the probability density $p(u_z)$ by a simple change of coordinate system, using the Jacobian determinant: 
\begin{equation}\label{eq:coord_transform}
    p(u_z) = p(z) \left|\frac{dz}{du_z}\right|
.\end{equation}
In a similar way, the radial probability density can first be expressed in Polar coordinates, and then transformed into the probability density $p(u_\rho)$ by another change of coordinates. 

Eventually, we can calculate the probability density of the combined field envelopes and thus the individual Purcell factor by integration:
\begin{equation}
    p(P_\text{Er}) = \int du_\rho\, p(u_\rho)\int du_z\, p(u_z) \, \delta(P_\text{Er}-P_\text{Er,max}u_z u_\rho)
,\end{equation}

which leads to the closed expression

\begin{equation}\label{eq:p_Er}
    p(P_\text{Er}) = \frac{\arctan(\sqrt{P_\text{Er,max}/P_\text{Er}-1})}{4\pi P_\text{Er}}.
\end{equation}

The resulting probability density is shown in Fig.~\ref{fig:cavity model}b as blue dotted line, which agrees with the results of our Monte Carlo simulation. 

Now, the effect of a specific Purcell enhancement factor on the contribution to the fluorescence signal is twofold \cite{reiserer_cavity-based_2015}: First, the probability to collect a photon emitted by a dopant into the cavity quickly drops with lower $P_\text{Er}$, as $p_\text{coll}(P_\text{Er})=P_\text{Er}/(P_\text{Er}+1)$  (black dotted curve). Second, also the probability $p_\text{ex}(P_\text{Er})$ to excite a given dopant within a time $\tau$ by an intracavity field with $n_\text{cav}$ photons depends on the coupling strength $g= \sqrt{P_\text{Er} \gamma\kappa}$ and thus the Purcell factor. Assuming coherent driving, i.e. $\tau \ll T_2$, we have: $p_\text{ex}(P_\text{Er})=\sin^2(\sqrt{n_\text{cav}} g \tau/2)$ (black dashed curve).

These two effects increase the relative contribution of ions with larger Purcell factor to the fluorescence signal, shown as effective density in Fig.\,\ref{fig:dopantdensity} (blue solid line). In this model, we find that the average Purcell enhancement in our experiment is less than half of the maximal value (red dashed line).

To predict the fluorescence decay curves, we numerically integrate the single-exponential curves for a random distribution of dopants within the resonator mode, considering the effects mentioned above:

\begin{equation}
    I(t) = \int\limits_1^{P_\text{Er,max}} dP_\text{Er}\;\, p(P_\text{Er})\,p_\text{coll}(P_\text{Er})\,p_\text{ex}(P_\text{Er})\;\text{e}^{-(P_\text{Er}+1)\frac{t}{T_0}}
\end{equation}

Here $T_0$ is the free space decay constant. The choice of a lower integration bound, $P_\text{Er} \geq 1$, is justified as only dopants with considerable Purcell enhancement contribute to our experimental signal, as detailed above. The obtained curve is in excellent agreement with the data, as shown in Fig.~2 of the main paper.

To accurately model the fluorescence curves in presence of cavity vibrations, we further include a reduction of the maximum Purcell factor according to the rms cavity detuning $\Delta\nu_\text{rms}$ \cite{reiserer_cavity-based_2015}:
\begin{equation}
    P_\text{Er,max}\sim \frac{1}{1+(\frac{\Delta\nu_\text{rms}}{\Delta\nu_\text{FWHM}/2})^2}
\end{equation}
Leaving this detuning as a fit parameter gives the inset of Fig.~2 of the main manuscript that shows the obtained cavity resonance stability as a function of lock laser power.

\subsection{Two- and three-pulse echo measurements}

In two-pulse experiments a first optical pulse with an area of approximately $\pi/2$ prepares an optical superposition state that quickly dephases because of the inhomogeneous broadening. A subsequent optical $\pi$-pulse, irradiated after a time $t/2$, leads to a rephasing and thus to the emission of an echo signal at time $t$. The echo area $A$ decays exponentially with the time $t$ spent in the superposition state \cite{bottger_optical_2006}:
\begin{equation}
    A(t) = A_0 \, \exp(-2t/T_2)
.\end{equation}

\begin{figure}
\includegraphics[width=1.0\columnwidth]{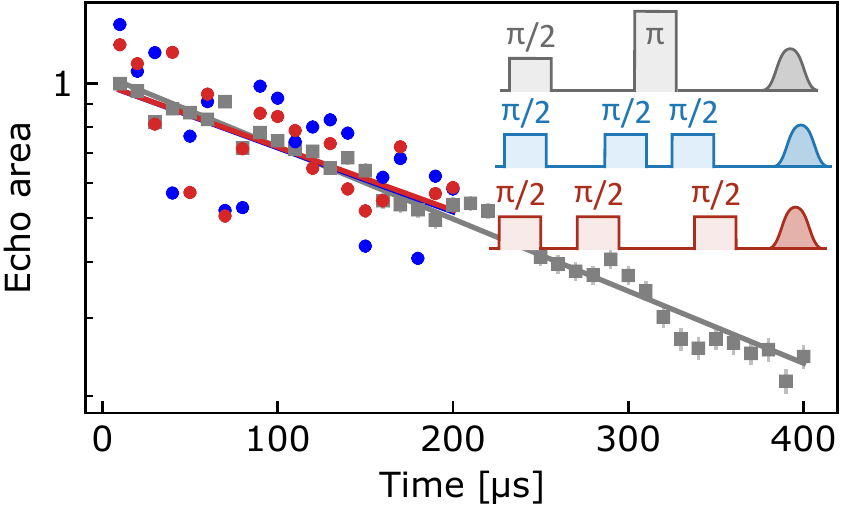}
\caption{\label{fig:echo}
\textbf{Two- and three-pulse echo decays.} In stimulated photon echo experiments, the echo area decays exponentially with time spent in the superposition state, indicating absence of spectral diffusion on that time scale (grey squares and line fit). The echo decay in three-pulse experiments show no difference between a waiting time of $T_w=4\,\si{\micro s}$ (blue circles and exponential fit) and $T_w=128\,\si{\micro s}$ (red) and exhibit the same decay constant as two-pulse measurements. 
}
\end{figure}

\begin{figure}[h!]
\includegraphics[width=1.0\columnwidth]{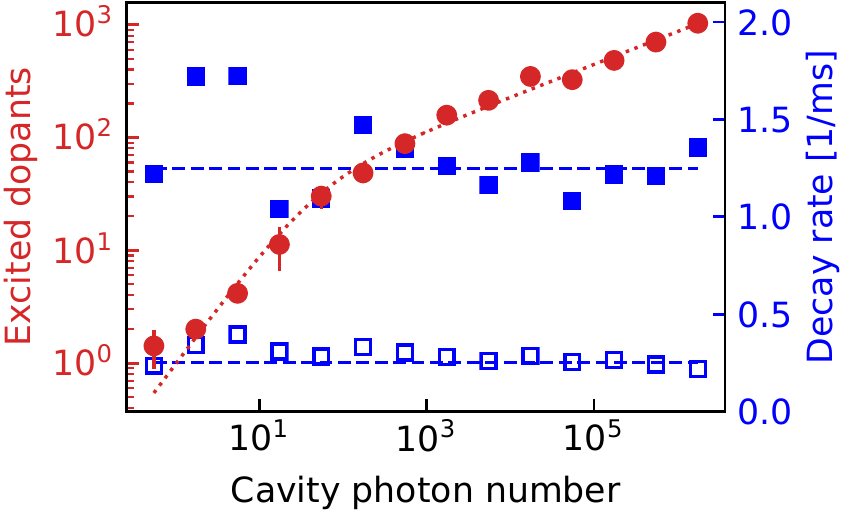}
\caption{\label{fig:dopantdensity}
\textbf{Power dependence of the fluorescence signal.} In fluorescence measurements with $500\,\si{\micro\second}$ long excitation pulses, we scan the resonant laser power. From biexponential fits we extract the fast (filled blue squares) and slow (open blue squares) decay rates, while we calculate the number of excited dopants from the average number of photon detection events (red points). The red dashed line is a fit to a model for a saturation curve with power broadening. 
}
\end{figure}

In three-pulse experiments, the refocusing $\pi$ pulse is replaced with a pair of $\pi/2$ pulses: the first one maps the superposition state onto the population in ground and excited state, while the second one restores the superposition and leads to rephasing. While the population grating during the waiting time between the pair of $\pi/2$ pulses is not subject to decoherence, the rephasing after the last pulse is sensitive to spectral diffusion during the waiting time. In general, the echo area for a time $t$ spent in the superposition state and a waiting time $T_w$ is given by \cite{bottger_optical_2006}
\begin{equation}
    A(t, T_w) = A_0 \, \exp\left(-\frac{2T_w}{T_1}\right)\,\exp\left(-2t\,\pi\Gamma_\text{eff}(t, T_w)\right)
.\end{equation}
For a fixed waiting time $T_w$, the echo decay now depends on the effective transition linewidth $\Gamma_\text{eff}$. The lower bound is given by the homogeneous linewidth $1/(\pi T_2)$, but in presence of spectral diffusion the effective linewidth will grow over time:
\begin{equation}
    \Gamma_\text{eff}(t, T_w) = \frac{1}{\pi T_2} + \frac{\Gamma_\text{SD}}{2}\left[\frac{Rt}{2}+\left(1-\exp(-RT_w)\right)\right]
\end{equation}
Here, $\Gamma_\text{SD}$ is the spectral width of the diffusion process, and $R$ is the rate at which it takes place. As shown in Fig. \ref{fig:echo}, we observe an exponential (rather than Gaussian) decay in both two- (grey) and three-pulse (blue and red) echo measurements. In addition, the effective linewidth, obtained by fitting the measured decay curves, stays constant (as also shown in the main text). We can thus conclude that $\sqrt{\Gamma_\text{SD} R}\ll 1/T_2$, which means that spectral diffusion is negligible on the investigated timescales. 

\subsection{Power dependence}
To study the effect of a varying intracavity photon number and determine the number of dopants coupled to the resonator, we vary the power in the excitation pulse. Fig.~\ref{fig:dopantdensity} shows that the decay constants (blue) of a biexponential fit performed for each power setting are constant. However, the number of excited dopants, calculated from the number of fluorescence photons and the independently determined outcoupling and detection efficiencies, increases with power (red). The nonlinear rise is explained by saturation effects, from which we can estimate the number of excited dopants before entering the power broadening regime to be about $50$. 
At this power level, the spectral width of the excitation will be comparable to the average single-dopant Rabi frequency $\Omega_\text{avg}$, which we calculate from the Purcell factor distribution $p(P_\text{Er})$ and the cavity photon number $n_\text{cav}\approx50$: $\Omega_\text{avg}=\braket{\sqrt{n_\text{cav} P_\text{Er} \kappa \gamma}}=2\pi\cdot 180\,\si{\kilo\hertz}$. 
Because of the intensity profile of the cavity mode, about half of the dopants in the crystal remain undetected due to their low Purcell factor. Therefore, the peak spectral density of a single magnetic class of dopants is about $0.56/\si{\kilo\hertz}$. 
By integration over the inhomogeneous linewidth, $\Gamma_\text{inh}=414(7)\,\si{\mega\hertz}$, correcting for the Boltzmann distribution between the ground states at finite temperature, and extrapolating the value to all four distinguishable dopant positions (two sites with two magnetic classes each), we arrive at an estimate of $10^7$ dopants in our cavity mode, corresponding to a density of $5.2\cdot 10^{15}\,\mathrm{cm}^{-3}$. Comparing this with the density of yttrium sites, $1.8\cdot 10^{22}\,\mathrm{cm}^{-3}$, we find a dopant concentration of $0.28\,\mathrm{ppm}$, in agreement with other measurements of undoped YSO from the same supplier \cite{dibos_atomic_2018}.

\bibliographystyle{naturemagsr.bst}
\bibliography{bibliography.bib}

 \end{document}